\documentclass[12pt]{iopart}
\usepackage{graphicx}
\usepackage{subfigure}
\usepackage{times}

\newcommand{\M}{\mathrm{M}}

\newcommand{\C}{\mathrm{C}}
\newcommand{\D}{\mathrm{D}}

\newcommand{\RR}{\mathrm{R}}

\newcommand{\ccr}{\hat{c}^\dagger}
\newcommand{\cde}{\hat{c}}
\newcommand{\acr}{\hat{a}^\dagger}
\newcommand{\ade}{\hat{a}}

\newcommand{\zetahat}{\hat{\zeta}}
\newcommand{\xihat}{\hat{\xi}}

\newcommand{\du}{\rmd}
\newcommand{\im}{\rmi}
\newcommand{\ex}{\rme}

\newcommand{\therm}{\mathrm{th}}

\newcommand{\eqref}[1]{(\ref{#1})}
\newcommand{\bfu}{\mathbf{u}}
\newcommand{\bfr}{\mathbf{r}}

\begin{document}

\title[Optomechanics with a high-frequency dilational mode in thin dielectric membranes]
{Quantum optomechanics with a high-frequency dilational mode in thin dielectric membranes}
\author{K. B{\o}rkje and S. M. Girvin}
%\author{J. G. E. Harris}
%\author{S. M. Girvin}
%\affiliation{Departments of Physics and Applied Physics, Yale University, New Haven, Connecticut 06520, USA}
\address{Departments of Physics and Applied Physics, Yale University, New Haven, Connecticut 06520, USA}

\date{Received \today}
\begin{abstract}
The interaction between a high-frequency dilational mode of a thin dielectric film and an optical cavity field is studied theoretically in the membrane-in-the-middle setup. A derivation from first principles leads to a multi-mode optomechanical Hamiltonian where multiple cavity modes are coupled by the thickness variation of the membrane. For membrane thicknesses on the order of one micron, the frequency of this dilational mode is in the GHz range. This can be matched to the free spectral range of the optical cavity, such that the mechanical oscillator will resonantly couple cavity modes at different frequencies. Furthermore, such a large mechanical frequency also means that the quantum ground state of motion can be reached with conventional refrigeration techniques. Estimation of the coupling strength with realistic parameters suggests that optomechanical effects can be observable with this dilational mode. It is shown how this system can be used as a quantum limited optical amplifier. The dilational motion can also lead to quantum correlations between cavity modes at different frequencies, which is quantified with an experimentally accessible two-mode squeezing spectrum. Finally, an explicit signature of radiation pressure shot noise in this system is identified.
\end{abstract}
%\pacs{42.50.-p, 42.65.-k, 03.65.Ta, 03.67.Bg}
\maketitle

\section{Introduction}

Research on mechanical systems in the quantum regime has reached several milestones in the past few years. A micromechanical oscillator in the quantum ground state was first observed in an experiment by O'Connell {\it et al.}~\cite{OConnell2010Nature}, where the high-frequency dilational motion of a slab of dielectric material was coupled to a superconducting qubit. This enabled verifying that the mechanical oscillator was in the ground state, as well as the ability to controllably create single phonons. The very high mechanical resonance frequency of $6$~GHz ensured that the quantum ground state was reached simply by conventional refrigeration techniques. 

Quantum behaviour of mechanical oscillators has later been achieved in cavity opto- or electromechanical systems, where the mechanical motion is coupled to light or microwaves. Resolved sideband cooling of mechanical motion to the quantum ground state \cite{Wilson-Rae2007PRL,Marquardt2007PRL} was achieved in an experiment by Teufel {\it et al.}~on an electromechanical system \cite{Teufel2011Nature}. This technique has also been successfully employed in the optical regime with silicon nanoscale resonators \cite{Chan2011Nature}, where the quantization of mechanical motion has been directly measured \cite{Safavi-Naeini2012PRL}. In a setup coupling light to the collective motion of cold atoms, quantization of motion \cite{Brahms2011} and ponderomotive squeezing of the light \cite{Brooks2011} have been reported.

The optomechanical system of a thin silicon nitride membrane placed inside an optical Fabry-P\'{e}rot cavity \cite{Thompson2008Nature} has also attracted considerable attention recently \cite{Jayich2008NJP,Wilson2009PRL,Clerk2010PRL,Nunnenkamp2010PRA,Sankey2010NatPhys,Xuereb2011PRL,Taylor2011}. The flexural modes of the membrane give rise to the same physics as a cavity with a movable end mirror. In addition, one can achieve coupling of the light to the square of the mechanical displacement \cite{Thompson2008Nature,Sankey2010NatPhys}, which could enable nondemolition measurements of the phonon number \cite{Jayich2008NJP} or of phonon shot noise \cite{Clerk2010PRL}. The idea of using such a membrane as a transducer between optical and electrical degrees of freedom has been proposed \cite{Taylor2011}. Finally, the membrane's coupling to light was recently exploited in an experiment to effectively couple the motion of the membrane and the motion of cold atoms \cite{Camerer2011PRL}. 

While the membrane-in-the-middle setup is promising, laser cooling of the flexural modes to the ground state is a considerable challenge. One reason for this is the typically low resonance frequencies of these modes, usually in the range of 0.1-10 MHz, which corresponds to a very low effective temperature necessary to reach the ground state. However, in addition to the flexural modes, the membrane also experiences thickness oscillations, i.e.~a dilational mode similar to the motion studied in the experiment by O'Connell {\it et al.}~\cite{OConnell2010Nature}. If the membrane is very thin, the thickness oscillation has a much higher frequency than the flexural modes. The frequency is approximately given by $v_\mathrm{s}/(2d_0)$, where $v_\mathrm{s}$ is the speed of sound in the membrane and $d_0$ is the membrane thickness. Note that thin-film dilational modes of this type have been detected in free-standing silicon membranes by using laser pulses \cite{Hudert2009PRB,Bruchhausen2011PRL}. For silicon nitride membranes with thicknesses of a micrometer or less, the mechanical resonance frequencies will be on the order of GHz or tens of GHz. Hence, provided that the coupling between this dilational mode and light is sufficiently strong, the membrane-in-the-middle setup could also give rise to an optomechanical system where the mechanical oscillator is in the ground state already at dilution refrigeration temperatures. Another useful feature of the very high mechanical frequency is that it is of the same size as the free spectral range in standard optical cavities, so the dilational mode will effectively couple cavity modes at different frequencies. This type of frequency matching was recently exploited in a different type of optomechanics experiment \cite{Bahl2012NatPhys}. 

In this article we investigate the coupling between the thickness variation of a thin dielectric membrane and an optical cavity field. In Sections \ref{sec:OptCavModes} and \ref{sec:MechModes}, we characterize the uncoupled optical and mechanical modes, respectively. The coupling comes about both due to the membrane thickness variation itself (a boundary effect) as well as the fact that elastic strain in the membrane leads to a change in the electric permittivity (a bulk effect). In general, both effects must be taken into account when light interacts with the motion of dielectric objects \cite{Bahl2012NatPhys,Rakich2012PRX}. In Section \ref{sec:Hamiltonian}, we derive a second quantized Hamiltonian and quantify the coupling strengths between the optical cavity modes and the mechanical oscillation. By inserting realistic parameters, we show in Section \ref{sec:estimation} that the coupling can be strong enough for optomechanical effects to be observable, provided that the as yet unknown mechanical quality factor exceeds a certain value. In Section \ref{sec:Amplifier}, we show how this system can be used as a quantum limited optical amplifier, due to the vanishingly small thermal occupation of the mechanical mode. Finally we study how one can produce quantum correlations between different cavity modes with this setup in Section \ref{sec:Correlations}. We calculate a two-mode squeezing spectrum and find that one can achieve considerable squeezing for sufficiently large mechanical quality factors. In addition, we identify an explicit signature of radiation pressure shot noise. 

\section{Optical cavity modes}
\label{sec:OptCavModes}

\begin{figure}[htbp]
\begin{center} 
\includegraphics[width=.8\textwidth]{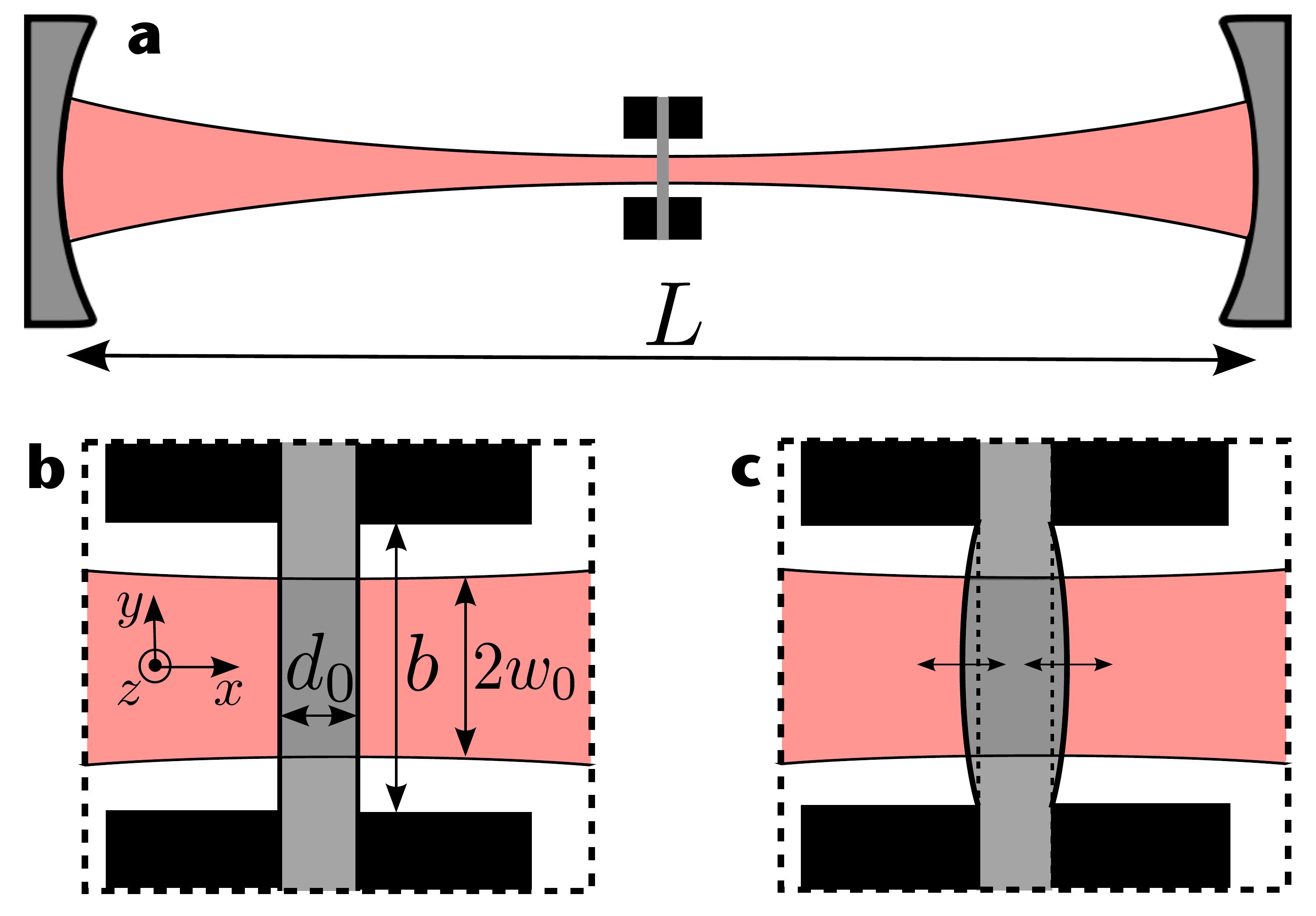}
\caption{Proposed experimental setup. {\bf a)} A thin dielectric membrane inside an optical Fabry-P\'{e}rot cavity. The membrane is placed close to the waist of the cavity mode. {\bf b)} Close-up view of the membrane when the membrane thickness is at its equilibrium value $d_0$. The waist radius of the cavity mode is $w_0$ and the side length of the square membrane is $b$. {\bf c)} Vibrational modes in the membrane leads to thickness fluctuations around the equilibrium value $d_0$.}
\label{fig:setup}
\end{center}
\end{figure}

We will consider the setup of a thin dielectric membrane placed inside an optical Fabry-P\'{e}rot cavity, depicted in Fig.~\ref{fig:setup}. The length of the cavity will be denoted by $L$, and the equilibrium membrane thickness by $d_0$. The center of the cavity is placed at $x=0$, and the right (left) mirror is positioned at $x =  \pm L/2$. The electric field inside the cavity can be expressed in terms of orthonormal modes $\mathbf{\Phi}_l(\mathbf{r})$,
\begin{equation}
  \label{eq:Efield}
  \hat{\mathbf{E}}(\mathbf{r}) = \sum_{l} \sqrt{\frac{\hbar \omega_{l}}{2 \varepsilon_0}} \left(\mathbf{\Phi}_{l}(\mathbf{r}) \hat{a}_{l} + \mathbf{\Phi}^\ast_{l}(\mathbf{r}) \hat{a}^\dagger_{l} \right) \ ,
\end{equation}
where $\omega_l$ is the mode frequency, $\varepsilon_0$ is the electric permittivity in vacuum, and $\hat{a}^{(\dagger)}_l$ are photon annihilation (creation) operators with $[\hat{a}_l,\hat{a}^\dagger_{l'}] = \delta_{l,l'}$. In the absence of the membrane, the functions $\mathbf{\Phi}_{l}(\mathbf{r})$ are the Hermite-Gauss modes \cite{Siegman1986Book}. In general, the subscript $l$ represents a set of quantum numbers. However, we will restrict ourselves to the Gaussian $\mathrm{TEM}_{00}$-modes only and consider polarized light. In that case, $l = 1,2,...$ will be the longitudinal mode index. 

The width of the cavity modes changes on the length scale of the Rayleigh range $x_\RR = k_l w_0^2/2$, where $k_l$ is the wavenumber and $w_0$ the waist size of the cavity mode. We will concentrate on the situation where the membrane is positioned at $x_\M$ well within the Rayleigh range from the waist, such that $|x_\M| \ll  x_\RR$. In this case, the mode functions inside the membrane can be approximated by $\mathbf{\Phi}_l(\mathbf{r}) = \Phi_l(\mathbf{r})  \mathbf{v}$ \cite{Biancofiore2011PRA} where $\mathbf{v}$ is a polarization vector of unit length and 
\begin{equation}
  \label{eq:uinmembrane}
  \Phi_l(\mathbf{r}) =  \frac{2 C_l}{\sqrt{\pi L} \, w_0} \ex^{-\left(\mathbf{r}_\parallel/w_0\right)^2} \sin (n_\M k_l x + \phi_l) \ .
\end{equation}
We have introduced $\mathbf{r}_\parallel = (y,z)$ as the component of the position vector in the $yz$-plane, and $n_\M$ is the membrane's index of refraction. The wave number is $k_l = \omega_l/c$, with $c$ being the speed of light. The dimensionless coefficients $C_l$ and the phases $\phi_l$ follow from normalization and boundary conditions. We do not include their expressions here, but note that $C_l$
%\begin{equation}
%  \label{eq:Cl}
%  C_l = \frac{1}{n_\M}\sqrt{(n_\M^2 - 1) \sin^2(k_l (x_\M - d_0/2 + L/2)) + 1} 
%\end{equation}
is of order 1 and $\phi_{l'} - \phi_l \approx (l' - l)\pi/2$ if the membrane's intensity reflection coefficient
\begin{equation}
  \label{eq:Reflcoeff}
%  R(k) = \frac{\left(n_\M^2 -1 \right)^2 \sin^2(n_\M k d_0)}{4 n_\M^2 \cos^2\left(n_\M k d_0\right) + \left(n_\M^2 + 1\right)^2 \sin^2(n_\M k d_0)} \ .
R(k) = \frac{\left(n_\M^2 -1 \right)^2 }{4 n_\M^2 \cot^2\left(n_\M k d_0\right) + \left(n_\M^2 + 1\right)^2 } \ 
\end{equation}
is much smaller than 1. For Si$_3$N$_4$, $n_\M \approx 2$, such that $R(k)$ oscillates between 0 and 0.36 as a function of $k d_0$.

\begin{figure}[htbp]
\begin{center} 
\includegraphics[width=.8\textwidth]{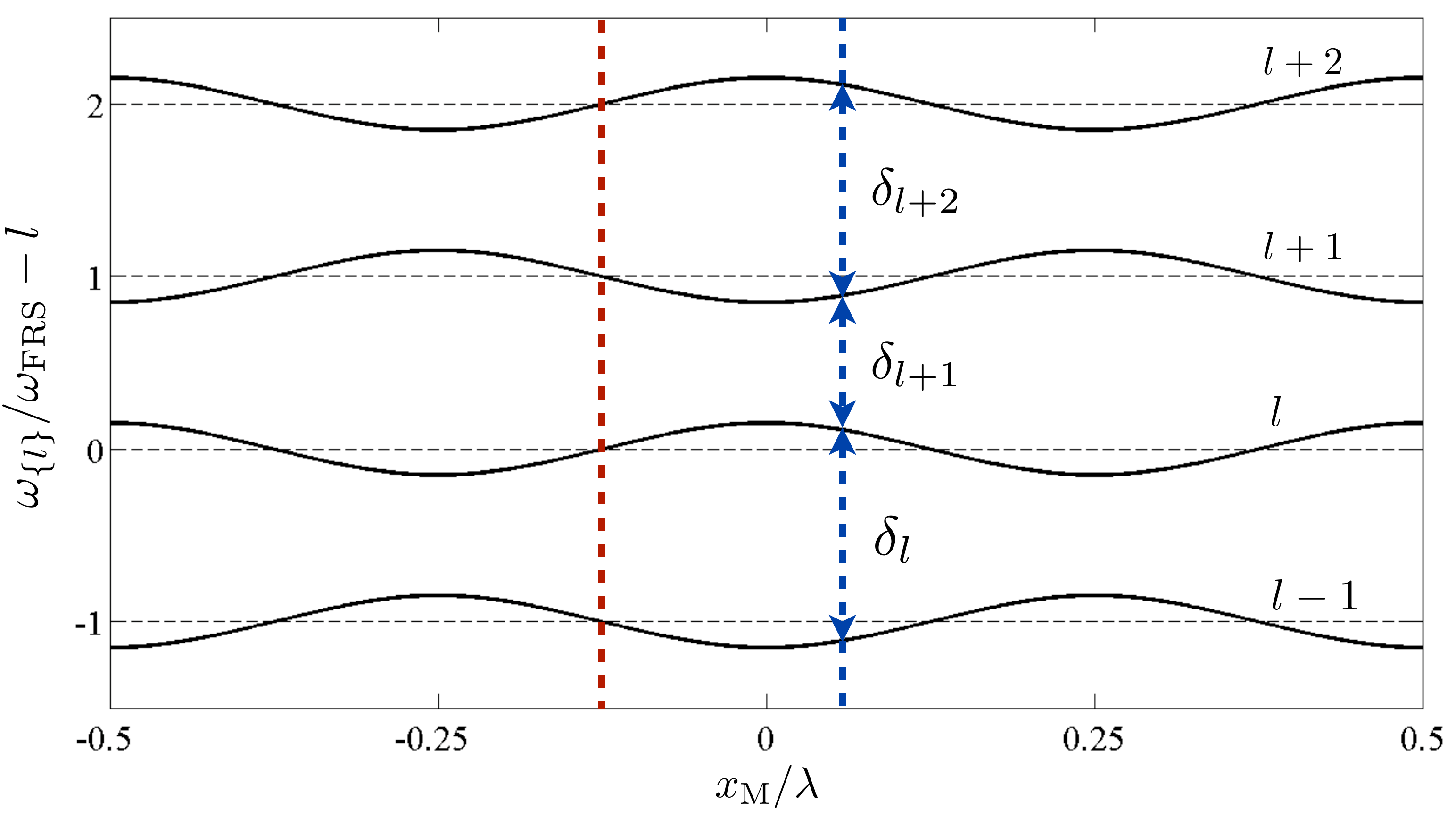}
\caption{Cavity mode resonance frequencies in units of $\omega_\mathrm{FSR} = c\pi/L$ as a function of membrane position $x_\M$. We have used a membrane thickness of $d_0 = 1 \ \mu$m, a wavelength of $\lambda = 1064$~nm, and an index of refraction $n_\M = 2$, giving $R(k) = 0.2$. The frequencies have an oscillatory dependence on $x_\M$, which means that they are in general not evenly spaced. The blue dashed line to the right is placed at a membrane position where the frequency spacings $\delta_l = \omega_l - \omega_{l-1}$ differ, i.e.~$\delta_l \neq \delta_{l+1} \neq \delta_{l + 2}$. The red dashed line to the left is placed at a high-symmetry point where $\delta_l = \delta_{l+1} = \delta_{l+2}$. }
\label{fig:freqs}
\end{center}
\end{figure}

The mode frequencies are given by $\omega_l = \omega^0_l + \delta \omega_l$, where $\omega^0_l = l \pi c/L$ are the evenly spaced frequencies in absence of the membrane for large longitudinal mode numbers $l$. The correction to the frequencies in presence of the membrane can be expressed in terms of $k^0_l \equiv \omega^0_l/c$ and becomes \cite{Biancofiore2011PRA}
\begin{equation}
  \label{eq:freqcorr}
  \delta \omega_l = \frac{c}{L} \left[\arcsin\left((-1)^l \sqrt{R(k^0_l)} \cos\left(2 k^0_l x_\M\right) \right) - \beta(k^0_l)  \right] \ .
\end{equation}
The frequencies have an oscillatory dependence on the position of the membrane \cite{Thompson2008Nature} whose magnitude depends on the reflection coefficient $R(k)$. Note also that the shifts for neighboring longitudinal numbers $l$ and $l + 1$ have opposite sign. The contribution $\beta(k^0_l)$ has only a weak dependence on $l$ and can be thought of as an unimportant shift of all frequencies. The mode frequencies as a function of membrane position are shown in Fig.~\ref{fig:freqs}.

\section{Dilational modes of the membrane}
\label{sec:MechModes}

In this section, we study the high-frequency dilational modes of the membrane, which is assumed to be square with side lengths $b$ and an equilibrium thickness $d_0$. The membrane lies in the $yz$-plane with its front and back surfaces at $x_{\pm} = x_\M \pm d_0/2$. The displacement field in the membrane will be denoted by $\bfu(\bfr)$. Assuming that the membrane material is isotropic, the displacement eigenmodes are given by the equation
\begin{equation}
  \label{eq:ElasticEOM}
  \mu \nabla^2 \bfu + \left(K + \frac{\mu}{3} \right) \nabla \left(\nabla \cdot \bfu\right) = - \rho \, \omega^2 \bfu ,
\end{equation}
where $K$ is the bulk modulus, $\mu$ is the shear modulus, $\rho$ is the mass density, and $\omega$ is the mode frequency. To find the displacement field, we also need to specify the boundary conditions, which depend on the details of how the membrane is connected to its mechanical support. For simplicity, we will work with the so-called simply supported boundary conditions
\begin{eqnarray}
  \label{eq:BoundCondElast}
  u_x = u_y = \sigma_{zz} = 0 \quad , \quad z = \pm b/2  \\
  u_x = u_z = \sigma_{yy} = 0 \quad , \quad y = \pm b/2 \nonumber \\
  u_y = u_z = \sigma_{xx} = 0 \quad , \quad x = x_\pm \nonumber 
\end{eqnarray}
where $\sigma_{ii}$ are the diagonal elements of the stress tensor. This means that the transverse displacement as well as the normal component of the stress on all surfaces are zero. This simplification might not be the most realistic choice of boundary conditions. However, we do not expect that other boundary conditions will bring significant changes to the properties we are interested in, namely the frequency spacings between the modes and the optomechanical couplings. 

The boundary conditions and the equation of motion are satisfied by the field $\bfu(\bfr) \propto \tilde{\bfu}_{mn}(\bfr) $ with
\begin{eqnarray}
  \label{eq:SolDisplacement}
  \tilde{u}_{mn,x}(\bfr) & = & K_x \sin \left[\frac{\pi}{d_0} (x - x_\M)\right] \sin\left[ m \frac{\pi}{b} \left(y + \frac{b}{2}\right) \right] \sin \left[n \frac{\pi}{b} \left(z + \frac{b}{2}\right) \right] \\
  \tilde{u}_{mn,y}(\bfr) & = & K_y \cos \left[\frac{\pi}{d_0} (x - x_\M)\right] \cos\left[ m \frac{\pi}{b} \left(y + \frac{b}{2}\right) \right] \sin \left[n \frac{\pi}{b} \left(z + \frac{b}{2}\right) \right] \nonumber \\
  \tilde{u}_{mn,z}(\bfr) & = & K_z \cos \left[\frac{\pi}{d_0} (x - x_\M)\right] \sin\left[ m \frac{\pi}{b} \left(y + \frac{b}{2}\right) \right] \cos \left[n \frac{\pi}{b} \left(z + \frac{b}{2}\right)\right] \nonumber \ , 
\end{eqnarray}
where $K_j$ are constants, $m,n = 1,2,3...$, and we have restricted ourselves to the lowest non-zero wave number ($\pi/d_0$) in the $x$-direction. The angular frequency $\omega = \omega_{mn}$ is given by
\begin{equation}
  \label{eq:FreqDilational}
  \omega_{mn} = v  \left[\left(\frac{\pi}{d_0} \right)^2 + \left(m^2 + n^2\right)\left(\frac{\pi}{b} \right)^2 \right]^\frac{1}{2}
\end{equation}
where the propagation velocity $v = v_\mathrm{s} \equiv \sqrt{(K + 4\mu/3)/\rho}$ for longitudinal waves and $v = \sqrt{\mu/\rho}$ for transversal waves.

Since the thickness $d_0$ is much smaller than the width $b$, the longitudinal vibrational modes will consist mostly of displacements in the $x$-direction, whereas the transverse modes consist mostly of displacements in the plane of the membrane. In the remainder of this article, we will focus on longitudinal modes only. We emphasize that the transverse modes can also couple to the optical cavity modes through the photoelastic effect. However, since the longitudinal and transverse mode frequencies do not coincide, we can safely ignore the latter below. The normalized longitudinal modes are given by Eq.~\eqref{eq:SolDisplacement} with $K_y/K_x = m d_0/b$, $K_z/K_x = n d_0/b$, and $K_x = (1/(1 + (m^2 + n^2) (d_0/b)^2))^{1/2} \approx 1$. Note that for these modes, both the normal and shear stress on the surfaces $x = x_\M \pm d_0/2$ are zero, as they should be on a free surface.

Ignoring the transverse modes, the total displacement field becomes $\bfu(\bfr) = \sum_{mn} \eta_{mn} \tilde{\bfu}_{mn}(\bfr)$. In other words, $\bfu(\bfr)$ is the displacement field of the vibrational modes with angular frequencies close to $\pi v_\mathrm{s}/d_0$. The length $\eta_{mn}$ characterizes the amplitude of the $mn$-mode. We can quantize the displacement field by letting $\eta_{mn} \rightarrow \eta_{\mathrm{zpf},mn} \left(\hat{c}_{mn} + \hat{c}^\dagger_{mn} \right)$, where $\hat{c}_{mn}$ is a phonon annihilation operator for the $mn$-mode. The zero point fluctuations are characterized by the length $\eta_{\mathrm{zpf},mn} = \sqrt{\hbar/(2 \tilde{m} \omega_{mn})}$, where $\tilde{m} = \rho d_0 b^2/8$ is the effective mass of each mode. 

We would like to study the coupling between the optical cavity modes and a single vibrational mode, namely the fundamental $m = n = 1$ mode. This is possible if the modes are well separated in frequency, which requires that the mechanical quality factor $Q \gg (b/d_0)^2$. This is one of several reasons for using a membrane with a small area to reduce the aspect ratio $b/d_0$. If this requirement for the mechanical $Q$ is too strict, it is however also possible to focus solely on the fundamental mode if the light couples much more strongly to it than to the other longitudinal modes. We will address this issue in the next section.

\section{The Hamiltonian for the coupled system}
\label{sec:Hamiltonian}

We now derive the Hamiltonian for the coupled optomechanical system. We assume that we can focus on a single mechanical mode with $m = n = 1$. We let $\omega_{11} \rightarrow \omega_\M \approx v_\mathrm{s} \pi/d_0$, $\eta_{mn} \rightarrow \eta$, and $\tilde{\bfu}_{11} \rightarrow \tilde{\bfu}$, such that the displacement field is $\bfu(\bfr) = \eta \, \tilde{\bfu}(\bfr)$. We define $\mathbf{r}_\mathrm{l} = (x_-,\mathbf{r}_\parallel)$ and $\mathbf{r}_\mathrm{r} = (x_+,\mathbf{r}_\parallel)$, which describe the equilibrium left and right membrane surfaces, respectively. Note that the membrane thickness at in-plane position $\bfr_\parallel$ is given by $d(\bfr_\parallel) = d_0 + 2 u_x(\bfr_\mathrm{r})$. The displacement field will later be quantized as described in Section \ref{sec:MechModes}. The coupling to other mechanical modes will be commented on in Section \ref{sec:CouplingOtherModes}.

\subsection{Derivation of the full Hamiltonian}
\label{sec:FullHam}

For a static membrane, the Hamiltonian is $\hat{H}_0 = \sum_l \hbar \omega_l \acr_l \ade_l$. In the presence of membrane vibrations, the Hamiltonian becomes $\hat{H} = \hat{H}_0 + \hat{H}_1 + \hat{H}_2 + \hat{H}_\M$, where $\hat{H}_\M$ is the energy of the mechanical fluctuations. The term $\hat{H}_1$ is the energy change of the electromagnetic field due to the change in membrane thickness, whereas $\hat{H}_2$ is the energy change due to the photoelastic effect, i.e.~the fact that a strain in the membrane gives rise to a change in the electric permittivity tensor. Using perturbation theory \cite{Johnson2002PRE}, these terms can be expressed as $\hat{H}_i = 1/2 \int \du \mathbf{r}_\parallel \hat{h}_i(\mathbf{r}_\parallel)$ where
\begin{eqnarray}
\label{eq:H1H2next}
  \hat{h}_1(\mathbf{r}_\parallel) & = & \eta \, \tilde{u}_x(\mathbf{r}_\mathrm{r}) \big(\varepsilon_{\M,0} - \varepsilon_0\big) \left[\hat{\mathbf{E}}^2(\mathbf{r}_\mathrm{l},t) + \hat{\mathbf{E}}^2(\mathbf{r}_\mathrm{r},t) \right]  \\
   \hat{h}_2(\mathbf{r}_\parallel) & = & \eta \int_{l_0}^{r_0} \du x  \, \frac{\partial \varepsilon_\M(\bfr)}{\partial \eta}\Bigg|_{\eta = 0} \, \hat{\mathbf{E}}^2(\mathbf{r},t) \ . \nonumber
\end{eqnarray}
Note that the expression for the surface contribution $\hat{H}_1$ is only well-defined when the electric field is continuous across the boundary \cite{Johnson2002PRE} and the optical modes are non-degenerate, both of which are satisfied here. The strain-dependent permittivity in the membrane is denoted by $\varepsilon_\M(\bfr)$, and $\varepsilon_{\M,0} = \varepsilon_0 \sqrt{n_\M}$ is its equilibrium value. We have assumed that the amplitude $\eta$ is small compared to the wavelength of the light and the equilibrium thickness $d_0$. We also assume that the optical cavity modes are polarized in the $yz$-plane, such that $\varepsilon_{\M}(\bfr)$ is the in-plane membrane permittivity. Note that when $\partial \varepsilon_\M(\bfr)/\partial \eta|_{\eta = 0} < 0$, i.e.~when the electric permittivity decreases as the thickness increases and vice versa, the contributions $\hat{H}_1$ and $\hat{H}_2$ have opposite signs.

In the presence of strain described by the strain tensor $e_{ij}$, the deviation of the inverse electric permittivity tensor $\varepsilon^{-1}$ from its equilibrium value is given by $\delta (\varepsilon^{-1})_{ij} = \sum_{kl} p_{ijkl} e_{kl}/\varepsilon_0$, where $p_{ijkl}$ is the photoelastic tensor \cite{Biegelsen1974PRL}. Assuming that the membrane is isotropic in the $yz$-plane and that $\hat{x},\hat{y},\hat{z}$ point along the principal axes of the membrane material, we have $(\varepsilon^{-1})_{ij} = \varepsilon_{ij}^{-1} \delta_{ij}$ with $\varepsilon_{yy} = \varepsilon_{zz} \equiv \varepsilon_\M$. Neglecting the small displacement components $\tilde{u}_y$ and $\tilde{u}_z$, we then arrive at $\partial \varepsilon_\M(\bfr)/\partial \eta|_{\eta = 0} \approx - \varepsilon_{\M,0}^2 p/\varepsilon_0 \, \partial \tilde{u}_x(\bfr) / \partial x$, where $p = p_{yyxx}$ and $\tilde{u}_x(\bfr)$ is given by Eq.~\eqref{eq:SolDisplacement}.
%$p = p_{yyxx} - \nu_{xy} (p_{yyyy} + p_{yyzz})$ and $\nu_{xy}$ is the Poisson ratio. 
For SiO$_2$, $p_{yyxx} = 0.271$ \cite{Biegelsen1974PRL}. We are not aware of any measurements of the photoelastic coefficients for Si$_3$N$_4$, but it seems reasonable to assume that they are not very different for Si$_3$N$_4$ than for SiO$_2$. In any case, they are expected to be of the same order of magnitude. 
%Note that for positive $p$, the total contributions from the change in thickness $\hat{H}_1(t)$ and the change in permittivity $\hat{H}_2(t)$ are of opposite sign. 
%This is as expected, since an increase in thickness also reduces the index of refraction of the membrane.

We now quantize the displacement field by letting $\eta \rightarrow \eta_\mathrm{zpf} \left(\cde + \ccr\right)$, where $\cde$ and $\ccr$ are phonon annihilation and creation operators and $\eta_\mathrm{zpf} = \sqrt{\hbar/(2 \tilde{m} \omega_\M)}$ is the size of the zero point fluctuations. By inserting \eqref{eq:Efield} and \eqref{eq:SolDisplacement} and ignoring terms that do not conserve photon number, we arrive at the multi-mode optomechanical Hamiltonian 
\begin{equation}
  \label{eq:Hfinal}
  \hat{H}  = \sum_l \hbar \omega_l \hat{a}^\dagger_l \hat{a}_l + \hbar \omega_\M \ccr \cde +  \sum_{l,l'} \hbar \Gamma_{l,l'} \left(\cde + \ccr\right) \hat{a}^\dagger_l \hat{a}_{l'} \ .
\end{equation}
%The coupling constant between mode $l$ and $l + 1$ is
%\begin{eqnarray}
%  \label{eq:gll}
  %G_{l,l'} & = & \frac{\sqrt{\omega_l \omega_{l'}}}{L} C_l C_{l'} N x_\mathrm{zpf} \\ & & \times \sin \left(n_\M k_l x_\M + \phi_l\right) \sin \left(n_\M k_{l'} x_\M + \phi_{l'}\right) \nonumber
%   G_{l,l + 1} & = & \frac{\sqrt{\omega_l \omega_{l'}}}{2 L} C_l C_{l'} \eta_\mathrm{zpf} \Big[ \cos\left((l-l')\pi/2
%\end{eqnarray}
Below, we will focus on the intermode coupling $\Gamma_{l,l+1} = \Gamma_{l+1,l}$ between adjacent longitudinal optical modes. Defining $g_l = \omega_l \eta_\mathrm{zpf}/L$ and inserting the expressions for $C_l$ and $\phi_l$, we find $\Gamma_{l,l + 1}  =  g_l \Lambda_{l,l+1}$, where 
\begin{eqnarray}
  \label{eq:Gllp1}
\fl \qquad \Lambda_{l,l + 1}  & =  & \frac{I_{1,1}}{4}  \left(1 - \frac{n_\M^4 p \left(n_\M^2 - 1 \right)^{-1}}{1 - (2 n_\M k_l d_0/\pi)^2 } \right) \frac{\sqrt{R(k_l)}  \cot(n_\M k_l d_0) }{\left(1 - R(k_l) \cos^2(2 k_l x_\M)\right)^{1/2}}  \sin \left(2 k_l x_\M\right) \ .
% \fl \ \Lambda_{l,l + 1}  & =  & \frac{I_{1,1}}{4}  \left(1 + \frac{n_\M^4 p \left(n_\M^2 - 1 \right)^{-1}}{1 + (2 n_\M k_l d_0/\pi)^2 } \right) \sqrt{\frac{R(k_l) }{1 - R(k_l) \cos^2(2 k_l x_\M)}}  \cot(n_\M k_l d_0) \sin \left(2 k_l x_\M\right) 
 % \fl \ \Lambda_{l,l + 1}  & =  & \frac{I_{1,1}  C_l C_{l+1}}{2} \sin \left[2 \left(n_\M k_l x_\M + \phi_l \right)\right] \cos(n_\M k_l d_0)  \left(n_\M^2 - 1  + \frac{\pi^2 n_\M^4 p}{\pi^2 + (2 n_\M k_l d_0)^2} \right)  
\end{eqnarray}
%is a dimensionless number of order 1. 
The quantity $I_{1,1}$ is a transverse overlap integral which is defined in Eq.~\eqref{eq:Itr} below. Note that $\sqrt{R(k_l)} \cot(n_\M k_l d_0) \rightarrow 0$ when $\cot(n_\M k_l d_0) \rightarrow \pm \infty$, and that $\cot(n_\M k_l d_0)/(1-(2n_\M k_l d_0/\pi)^2) \rightarrow \pi/4$ when $n_\M k_l d_0 \rightarrow \pi/2$. Thus, except for special values of the membrane position $x_\M$ or the thickness $d_0$ for which the coupling vanishes, the dimensionless number $\Lambda_{l,l+1}$ is always of order 1. The coupling strength is therefore determined by $g_l$, which is the familiar coupling rate in the canonical optomechanical system of a cavity with a movable mirror. This also determines the size of the intramode coupling $\Gamma_{l,l}$. We have used $k_l x_\M \approx k_{l+1} x_\M$, $\sqrt{\omega_l \omega_{l+1}}\approx \omega_{l}$, and $\phi_{l+1} - \phi_l \approx \pi/2$ to simplify the expression for the coupling strength. These are very good approximations for large longitudinal numbers $l$ and small reflectivities $R(k_l)$. For further details on the derivation of Eq.~\eqref{eq:Gllp1}, we refer to Ref.~\cite{Biancofiore2011PRA}, where the calculation of coupling strengths to the flexural modes produce similar expressions. 

The first term in the paranthesis in Eq.~\eqref{eq:Gllp1} comes from the thickness variation whereas the second from the photoelastic effect. We see that these terms are of opposite signs for $d_0 < \pi/(2 n_\M k_l)$ and of the same sign for $d_0 > \pi/(2 n_\M k_l)$. This does not contradict the fact that $\hat{H}_1$ and $\hat{H}_2$ always have opposite signs for $p > 0$, since we are now considering the coupling between two particular optical modes, and not the entire Hamiltonian. We also see that the two terms are of comparable size for reasonable parameters. Finally, note that the photoelastic term vanishes when the membrane thickness becomes much larger than the optical wavelength. This is as expected, since this term is, roughly speaking, an integral of two orthogonal optical modes over the thickness of the membrane.

\subsection{Reduction to two optical modes}

When focusing on a single optical mode with index $l$ and the intramode coupling $\Gamma_{l,l}$ only, the Hamiltonian is reduced to that of a standard optomechanical system. However, for thin membranes, the frequency of the dilational mode $\omega_\M$ will typically be very large compared to a typical cavity linewidth, denoted below by $\kappa$. Indeed, the mechanical frequency can be matched to the free spectral range of the cavity. Hence, we will consider two optical modes $l$ and $l + 1$ that are separated by the mechanical frequency $\omega_\M$, such that the intermode terms proportional to $\Gamma_{l,l+1}$ and $\Gamma_{l+1,l}$ are resonant. This means that the annihilation (creation) of a phonon will convert a photon from mode $l$ ($l + 1$) to mode $l + 1$ ($l$). Defining $\delta_l = \omega_l - \omega_{l-1}$, note that the frequency shift \eqref{eq:freqcorr} ensures that $\delta_{l+1}  \neq \delta_l \approx \delta_{l+2}$ whenever $\cos(2 k_l x_\M) \approx \cos(2 k^0_l x_\M) \neq 0$ (see Fig.~\ref{fig:freqs}). Thus, by placing the membrane at a position where $|\delta_{l+1} - \delta_{l}| \gg \kappa$, we can focus on just two optical modes. Furthermore, in the resolved sideband regime $\omega_\M \gg \kappa$, we can neglect the intramode coupling $\Gamma_{l,l}$ as well as the counter-rotating terms in the intermode coupling. 

The maximal difference between $\delta_{l+1}$ and $\delta_l$ occurs when $\cos(2 k_l x_\M) = \pm 1$. However, according to Eq.~\eqref{eq:Gllp1}, the coupling rate $\Gamma_{l,l+1}$ is zero at these points. This is not expected to cause a problem in practice since maximizing $|\delta_{l+1} - \delta_l|$ is not really necessary. One only needs to ensure that $|\delta_{l+1} - \delta_l| \gg \kappa$. To show that this is feasible, let us for example assume that $\cos(2 k_l^0 x_\M) = 1/2$ and $\sqrt{R(k)} = 0.2$. By expressing the linewidth $\kappa = c\pi/(L {\cal F})$ by the cavity finesse ${\cal F}$, we get $|\delta_{l+1} - \delta_l|/\kappa \sim 0.1 \, {\cal F} \gg 1$ for any reasonable finesse. Hence, we can safely focus on two cavity modes and neglect the others. The vertical blue (dashed) line in Fig.~\ref{fig:freqs} is placed at a membrane position where both $\delta_{l + 1} \neq \delta_l, \delta_{l + 2}$ and the coupling rate $\Gamma_{l,l+1}$ is nonzero.

For a given $l$, we let $\hat{a}_l \rightarrow \hat{a}_1$, $\hat{a}_{l+1} \rightarrow \hat{a}_2$, and $\Gamma_{l,l+1} = \Gamma$, which leads to a model where two longitudinal cavity modes are coupled by the fundamental dilational mode of the membrane,
\begin{eqnarray}
  \label{eq:TwoModeModel}
   \hat{H}  = \sum_{i=1}^{2} \hbar \omega_i \hat{a}^\dagger_i \hat{a}_i + \hbar \omega_\M \ccr \cde + \hbar \Gamma \left(\cde \, \hat{a}^\dagger_2 \hat{a}_{1} + \ccr \hat{a}^\dagger_1 \hat{a}_{2} \right) \ ,
\end{eqnarray}
with $\omega_2 - \omega_1 \sim \omega_\M$. The indices 1 and 2 are just for labeling the cavity modes and do not represent the actual longitudinal numbers $l, l +1 \gg 1$. We note that the model \eqref{eq:TwoModeModel} was recently studied in the strong coupling regime $\Gamma \sim \kappa$ \cite{Ludwig2012,Stannigel2012}, where e.g.~quantum nondemolition measurements of photon and phonon numbers become possible. The coupling in the experimental setup we study here is however not strong enough to see these effects.

\subsection{Coupling to other mechanical modes}
\label{sec:CouplingOtherModes}

\begin{figure}[htbp]
\begin{center} 
\includegraphics[width=.8\textwidth]{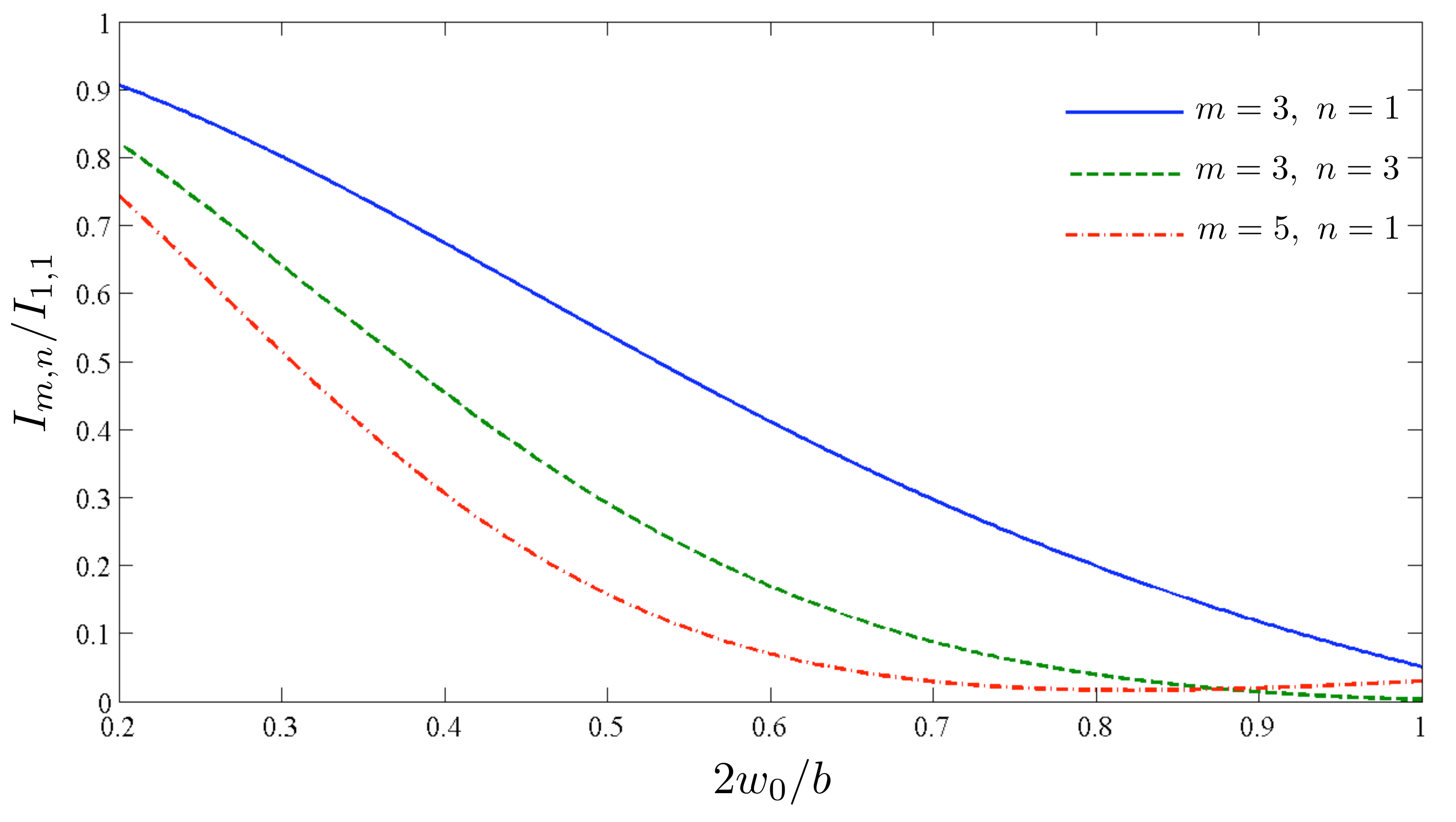}
\caption{The transverse overlap integral $I_{m,n}$ normalized to the integral for the fundamental mode $I_{1,1}$ as a function of the cavity waist diameter $2 w_0$. This shows that when the waist size becomes comparable to the membrane's transverse dimension $b$, coupling of the light to the fundamental mode $m = n = 1$ is significantly stronger than to the other vibrational modes.}
\label{fig:OtherMechModes}
\end{center}
\end{figure}

In the derivation of the interaction Hamiltonian we ignored all mechanical modes except the longitudinal fundamental mode with $m = n = 1$. However, as noted in Section \ref{sec:MechModes}, unless the mechanical quality factor $Q \gg (b/d_0)^2$, this can only be justified if the optomechanical coupling to other mechanical modes close in frequency is significantly smaller. In Eq.~\eqref{eq:Gllp1}, we expressed the coupling rate to the fundamental mode by the overlap integral $I_{1,1}$, defined as
\begin{equation}
  \label{eq:Itr}
  I_{1,1} = \frac{2}{\pi w_0^2} \int_A \du \bfr_\parallel \sin \left[\frac{\pi}{b} \left(y + \frac{b}{2}\right) \right]    \sin \left[\frac{\pi}{b} \left(z + \frac{b}{2}\right)\right] \,  \ex^{-2( \bfr_\parallel/w_0)^2} \ ,
\end{equation}
where the integral is over the transverse area of the membrane and we have assumed that the cavity waist is centered on the membrane. The integral $I_{1,1}$ is always of order 1. For other longitudinal mechanical modes that are close to the fundamental in frequency, i.e.~modes with arbitrary but moderate values of $m$ and $n$, the coupling $\Gamma_{l,l+1}$ will be given by the same expression as in Eq.~\eqref{eq:Gllp1}, except for the substitution $I_{1,1} \rightarrow I_{m,n}$, with
\begin{equation}
  \label{eq:Itrmn}
   I_{m,n} = \frac{2}{\pi w_0^2} \int_A \du \bfr_\parallel \sin \left[m \frac{\pi}{b} \left(y + \frac{b}{2}\right) \right]    \sin \left[n \frac{\pi}{b} \left(z + \frac{b}{2}\right)\right] \,  \ex^{-2( \bfr_\parallel/w_0)^2} \ .
\end{equation}
Symmetry dictates that the coupling to modes with $m$ or $n$ even becomes zero. In Fig.~\ref{fig:OtherMechModes}, we plot the overlap integral $I_{m,n}$ normalized to the value of $I_{1,1}$ for the three modes closest in frequency to the fundamental mode as a function of cavity waist size. 
We see that when the cavity waist is comparable in size to the membrane's transverse dimensions, i.e.~$2 w_0 \sim b$, the optomechanical coupling to the fundamental mode is dominant, which is another reason for using a membrane with a small side length $b$. The neglect of the other modes with $m$ or $n > 1$ thus appears to be justifiable, although one should keep in mind that they are there and might influence experiments.

\subsection{Strong optical drive}

In Section \ref{sec:estimation}, we will show that the coupling strength $\Gamma$ in Eq.~\eqref{eq:TwoModeModel} is typically very small compared to the cavity and oscillator decay rates. We therefore consider the situation where one of the cavity modes is strongly driven. Let us first assume that the upper mode (mode 2) is strongly driven by a laser with frequency $\omega_\D \sim \omega_2$. Transforming to the frame rotating at the drive frequency $\omega_\D$, and neglecting small nonlinear terms, we get the parametric amplification Hamiltonian
\begin{eqnarray}
  \label{eq:ParAmp}
  \hat{H} = - \hbar \Delta_1 \hat{a}^\dagger_1 \hat{a}_1 + \hbar \omega_\M \ccr \cde + \hbar \alpha_2 \left(\cde \,  \hat{a}_{1} + \ccr \hat{a}^\dagger_1 \right) \ .
\end{eqnarray}
Here, $\Delta_1 = \omega_\D - \omega_1 \sim \omega_\M$, and $\alpha_2 = \Gamma \sqrt{N_2}$, where $N_2$ is the average number of photons in cavity mode 2.
% \begin{equation}
%   \label{eq:alpha2}
%   \alpha_2 = G \sqrt{\frac{\kappa_\mathrm{ext} P}{\hbar \omega_\D\left[(\kappa/2)^2 + (\omega_\D - \omega_2)^2\right]}}
% \end{equation}
Alternatively, if mode 1 is driven at a frequency $\omega_\D \sim \omega_1$, we get the beam-splitter Hamiltonian
\begin{equation}
  \label{eq:beamSplitt}
  \hat{H} = - \hbar \Delta_2 \hat{a}^\dagger_2 \hat{a}_2 + \hbar \omega_\M \ccr \cde + \hbar \alpha_1 \left(\cde \,  \hat{a}^\dagger_{2} + \ccr \hat{a}_2 \right)\ ,
\end{equation}
in the frame rotating at $\omega_\D$ with $\Delta_2 = \omega_\D - \omega_2 \sim - \omega_\M$ and $\alpha_1 = \Gamma \sqrt{N_1}$ with $N_1$ the number of photons in cavity mode 1. It is well known that these Hamiltonians can be realized in the resolved sideband regime in ordinary single-mode cavity optomechanics by driving at either the blue or red sideband frequency. The difference in this case is that the drive frequencies are on resonance with other cavity modes, similar to the situation in Ref. \cite{Bahl2012NatPhys}.

Finally, we mention that one can also realize a beam-splitter Hamiltonian between the two cavity modes $\hat{a}_1$ and $\hat{a}_2$ by driving the mechanical oscillator. The mechanical motion can be induced either mechanically or by strongly driving two additional optical modes that are also separated by the mechanical frequency. This could allow for transfer of an optical signal from one frequency to another.

\section{Estimation of the coupling constant}
\label{sec:estimation}

In this section, we provide an estimate for the coupling strength between the dilational motion and the cavity modes. The size of the zero point fluctuations is independent of the membrane thickness for $b \gg d_0$ and is given by $\eta_\mathrm{zpf} \approx 2 \sqrt{\hbar/(\pi \rho b^2 v_s)}$. The speed of sound in Si$_3$N$_4$ is $v_s = 9900$~m/s and the density is $\rho = 3.44$~g/cm$^3$. Assuming that the membrane side length is $b =0.1$~mm, we get $\eta_\mathrm{zpf} = 2.0 \times 10^{-17}$~m. A smaller $b$ will reduce the membrane mass, which gives a larger zero point motion and hence a larger optomechanical coupling. This is yet another argument for using a small membrane.

We will imagine that we are using laser light with a wavelength of $\lambda = 1064$~nm. For membrane thicknesses of $d_0 = 100$~nm and $d_0 = 1 \ \mu$m, the mechanical frequency is $\omega_\M = 2\pi \times 50$~GHz and $\omega_\M = 2\pi \times 5$~GHz, respectively. This is relatively large compared to many other optomechanical realizations, although some experiments feature mechanical frequencies in the GHz regime \cite{Safavi-Naeini2011Nature,Carmon2007PRL,Ding2011ApplPhysLett}. The large resonance frequency means that at dilution refrigeration temperatures, the thermal phonon occupation number 
\begin{equation}
  \label{eq:ntherm}
  n_\therm = \frac{1}{\mathrm{exp}(\hbar \omega_\M/k_\mathrm{B} T) - 1}
\end{equation}
will be vanishingly small ($n_\therm < 10^{-5}$ even for $\omega_\M = 2\pi \times 5$~GHz), such that the mechanical oscillator will practically be in the quantum ground state without the need for additional laser cooling.

In Section \ref{sec:FullHam}, we saw that the optomechanical coupling strength is determined by the rate $g/2 \pi = c \, \eta_\mathrm{zpf}/(\lambda \, L)$. For a cavity of length $L = 1$~cm, this becomes $0.7$~Hz. This is comparable to the coupling strength to the flexural modes of the membrane in experiments performed with longer cavities \cite{Thompson2008Nature}. Naively, this suggests that optomechanical effects are observable also with the dilational mode. The mechanical linewidth $\gamma$ is however expected to be much larger for the high frequency dilational mode than for the low frequency flexural modes. What we need to check is how the effective coupling rate $\alpha \sim g \sqrt{N}$ compares to the cavity decay rate $\kappa$ and the mechanical oscillator decay rate $\gamma$. Here, $N$ is the average number of photons in the driven cavity mode. A figure of merit is the cooperativity $C = 4 \alpha^2/(\kappa \gamma)$. Introducing the mechanical quality factor $Q = \omega_\M/\gamma$ and the finesse ${\cal F} = c\pi/(L \kappa)$, we arrive at
\begin{equation}
  \label{eq:Coop}
   %C = \frac{192 \, d_0 P_\mathrm{in} {\cal F}^2 Q}{\pi^3 \lambda \, \rho A \, c \, v_s^2} \ .
   C \sim \left(\frac{4\eta_\mathrm{zpf}}{\lambda}\right)^2 \frac{c \, d_0}{v_s L} \, N {\cal F} Q \ 
\end{equation}
as an order of magnitude estimate for the cooperativity.

Inserting the expression for $\eta_\mathrm{zpf}$, and expressing the average photon number $N= 4P_\mathrm{in} \lambda/(h  c \,  \kappa)$ in terms of the laser power $P_\mathrm{in}$, our estimate for the cooperativity becomes
\begin{equation}
  \label{eq:Coop2}
  C \sim  \frac{128 \, d_0 P_\mathrm{in} {\cal F}^2 Q}{\pi^3 \lambda \, \rho b^2 \, c \, v_s^2} \ .
\end{equation}
The only unknown in this estimate is the mechanical $Q$. This can be experimentally determined by for example driving the mechanical motion with two laser beams separated by the mechanical frequency and analyzing the response of the oscillator. Assuming $d_0 = 1 \ \mu$m, $P_\mathrm{in} = 1.5$~mW, and ${\cal F}$ = 10$^5$, we get $C/Q \sim 5.8 \times 10^{-5}$. This means that for a mechanical $Q > 10^4$, the cooperativity can be on the order of 1, such that optomechanical effects should be observable. 

To realize the model \eqref{eq:TwoModeModel}, we want the frequency difference between the two optical modes to equal the mechanical frequency, i.e.~$\delta_{l+1} = \omega_\M$. In the absence of the membrane, the free spectral range of the cavity is $\delta_l = \pi c/L$. Thus, the length of the optical cavity will be determined by the requirement $L/d_0 \sim c/v_s$. Let us assume a membrane thickness of $d_0 = 1 \ \mu$m, which means that the cavity will need to be approximately $L = 3$~cm. In practice, since the difference $\delta_l$ depends on the membrane position $x_\M$ (see Fig.~\ref{fig:freqs}), one can move the membrane to the appropriate position where $\delta_l$ exactly matches $\omega_\M$. Note that with this frequency matching, the resolved sideband parameter $\omega_\M/\kappa$ is roughly equal to the finesse ${\cal F}$ of the optical cavity. 

With the assumptions above, the cavity linewidth is $\kappa = 2\pi \times 50$~kHz and $\alpha \sim 2 \pi \times 60$~kHz. For a mechanical quality factor of $Q \sim 10^4$, we have $\gamma = 2\pi \times 0.5$~MHz. Thus, even for modest cooperativity, since the effective coupling $\alpha$ exceeds $\kappa$, one could be able to observe strong coupling phenomena such as normal mode splitting of the mechanical resonance \cite{Groblacher2009Nature,Teufel2011Nature_2}. Significant resolved sideband cooling of the mechanical motion requires $C \gg 1$, which, with the numbers we have assumed here, is only possible if $Q \gg 10^4$. However, as noted earlier, the thermal occupation will be vanishingly small at millikelvin temperatures.

It is difficult to give an estimate for what the mechanical quality factor $Q$ of the dilational mode will be. It is however worth mentioning that extremely large $Q$ factors exceeding $10^8$ have recently been realized in similar structures, by making one of the surfaces convex, so as to provide confinement of the acoustic wave \cite{Goryachev2012}. One might imagine that a similar technique can be used to engineer large quality factors in the thin membranes we consider here, although one should note that the confinement reported in Ref.~\cite{Goryachev2012} is best for higher overtones of the fundamental mechanical resonance. For such overtones, where the mechanical frequency is $\omega_\M \sim q \, \pi v_s/d_0$ with $q > 1$ being an integer, the photoelastic term ($\hat{H}_2$) will give a negligible contribution to the optomechanical coupling. The surface term ($\hat{H}_1$) could however give rise to an optomechanical coupling for odd $q$. The coupling will still be reduced by a factor $1/\sqrt{q}$ compared to the fundamental mechanical mode, due to smaller zero point fluctuations. To confine the fundamental $q = 1$ mode and avoid propagation losses, it would be preferable to use a membrane whose thickness is sharply reduced at the edges.

Another issue one might worry about is optical loss in the membrane. Note however that membrane loss could not be observed in experiments with 50 nm thick membranes and a cavity finesse of $5 \times 10^4$ \cite{Sankey2010NatPhys}. While this does not exclude the possibility that it could be a problem at a thickness of 1 $\mu$m, it is worth pointing out that some of the absorption and scattering in the membrane is almost certainly due to surface roughness and contamination, which should not scale at all with thickness.

\section{Quantum limited optical amplification}
\label{sec:Amplifier}

We will now show how the optomechanical coupling to the dilational mode of the membrane can be used for quantum limited optical amplification. We consider the situation where a strong laser at frequency $\omega_\D$ drives the upper of two cavity modes separated by the mechanical frequency, which gives rise to the effective Hamiltonian \eqref{eq:ParAmp}. The detuning between the drive and the lower cavity resonance will be denoted $\Delta_1 = \omega_\D - \omega_1 \approx \omega_\M$. In addition to the strong drive, we imagine that a narrow bandwidth signal centered on $\omega_1$ is sent to the cavity, represented by the operator $\ex^{-\im \omega_1 t} \ade_{1,\mathrm{in}}(t)$. The bandwidth $B$ should be smaller than the decay rates $\kappa, \gamma$, which for the parameters in Section~\ref{sec:estimation} means roughly $B < 50$~kHz. The decay rate of the mirror through which the laser couples to the cavity modes is denoted by $\kappa_\mathrm{ex}$. We will account for other decay channels, e.g.~leakage through the other mirror or absorptive losses in the mirrors, whose respective decay rates are $\kappa_j$ with $j$ integer, such that $\kappa = \kappa_\mathrm{ex} +\sum_j \kappa_j$. In a frame rotating at $\omega_1$, input-output theory \cite{Collett1984PRA,Clerk2010RMP} gives the quantum Langevin equations
\begin{eqnarray}
  \label{eq:quantLang}
  \dot{\ade}_1 & = & -\frac{\kappa}{2} \ade_1 - \im \alpha_2 \, \ex^{-\im \Delta_1 t} \, \ccr + \sqrt{\kappa_\mathrm{ex}} \ade_{1,\mathrm{in}} + \sum_j \sqrt{\kappa_j} \, \xihat_j  \nonumber \\ 
  \dot{\cde} & = & - \left(\frac{\gamma}{2} + \im \omega_\M \right) \cde - \im \alpha_2 \, \ex^{-\im \Delta_1 t}  \, \acr_1 + \sqrt{\gamma} \, \zetahat \ .
\end{eqnarray}
Here, the operator $\zetahat(t)$ represents the noise from the mechanical bath. For a large mechanical $Q$, we can use a white noise model with $\langle \zetahat^\dagger(t) \zetahat(t') \rangle = n_\therm \delta(t-t')$ and $\langle \zetahat(t) \zetahat^\dagger(t') \rangle = \left(n_\therm + 1\right) \delta(t-t')$, where $n_\therm$ is the thermal phonon number defined in Eq.~\eqref{eq:ntherm}. The operators $\xihat_j$ represent the vacuum noise of the electromagnetic field and obey $\langle \xihat^\dagger_j(t) \xihat_k(t') \rangle = 0$ and $\langle \xihat_j(t) \xihat^\dagger_k(t') \rangle = \delta_{jk} \delta(t-t')$. The output field of cavity mode 1 is $\ade_{1,\mathrm{out}}(t) = \sqrt{\kappa_\mathrm{ex}} \ade_1(t) - \ade_{1,\mathrm{in}}(t)$ in a frame rotating at $\omega_1$. 

We define the Fourier transform as $\hat{f}^{(\dagger)}[\omega] = \int_{-\infty}^\infty \du t \, \ex^{\im \omega t} \hat{f}^{(\dagger)}(t)$. For frequencies $\omega \ll \kappa , \gamma$, the output field is
\begin{eqnarray}
  \label{eq:solution}
\ade_{1,\mathrm{out}}[\omega] & = & \frac{(2 r - 1 + C) \ade_{1,\mathrm{in}}[\omega] + 2\sqrt{r} \, \sum_j \sqrt{\kappa_j/\kappa} \, \xihat_j[\omega] - \im \sqrt{C} \, \zetahat^\dagger[\omega]}{1 - C} \ ,
\end{eqnarray}
where $r = \kappa_\mathrm{ex}/\kappa$ and $C = 4 \alpha_2^2/(\kappa \gamma)$ is the cooperativity. The result is not valid for $C \geq 1$, where nonlinear terms must be taken into account and self-induced mechanical oscillations occur \cite{Metzger2008PRL}. In the regime $1 - r < C < 1$, the interaction with the cavity leads to a phase-preserving amplified signal. Such an optomechanical amplifier has been realized in the microwave regime \cite{Massel2011}, although the quantum limit of added noise was not reached and the bandwidth was only on the order of kHz. The high-frequency dilational mode discussed here has the advantages that the thermal occupation $n_\therm$ can be made vanishingly small and the bandwidth can be significantly larger, due to a presumably larger mechanical linewidth. 

Defining the arbitrary quadrature $\hat{X}_{\theta, \mathrm{in}} = (\ex^{-\im \theta} \ade_{1,\mathrm{in}} + \ex^{\im \theta} \ade^\dagger_{1,\mathrm{in}})/\sqrt{2}$, we get 
\begin{equation}
  \label{eq:Xout}
  \langle \hat{X}^{2}_{\theta, \mathrm{out}} \rangle = G \Big(\langle \hat{X}^{2}_{\theta, \mathrm{in}} \rangle + (\Delta X)^2 \Big) \ ,
\end{equation}
where the amplifier gain is
\begin{equation}
  \label{eq:gain}
  G = \left(\frac{2 r - 1 + C}{1 - C}\right)^2 
\end{equation}
and the added noise is
\begin{equation}
  \label{eq:Noise}
 (\Delta X)^2 = \frac{1}{2} \left(1 - \frac{1}{G} \right) + \left(\frac{1}{r} - 1 \right) \left(1 + \frac{1}{\sqrt{G}}\right)^2   
  % (\Delta X)^2 = \frac{2 r (1-r) + 4 r \, C (n_\therm + 1/2)}{\left(2 r - 1 + C\right)^2} \ .
\end{equation}
when $n_\therm \ll 1$. If other types of cavity decay can be neglected, i.e. if $\kappa_\mathrm{ex} \approx \kappa$, we have $(\Delta X)^2 = (1 - 1/G)/2$ and the amplifier is quantum limited \cite{Caves1982PRD}. In the general case and for large gain $G \gg 1$, the added noise becomes $(\Delta X)^2 = 1/r - 1/2 \geq 1/2$. This means that in practice, deviations from the quantum limit will be due to amplification of vacuum noise entering the cavity through other decay channels, and not due to the mechanical oscillator.

Note also that for cooperativities $C < 1 - r$, the interaction with the cavity actually leads to absorption, in the sense that the gain $G$ is less than unity. In this regime, the magnitude of the reflected signal can even become smaller than for no optomechanical coupling ($C = 0$) \cite{Safavi-Naeini2011Nature}, and a portion of the signal is lost to the other optical decay channels. This has been observed in an optomechanical system \cite{Safavi-Naeini2011Nature}, and is analogous to electromagnetically induced absorption in atomic gases \cite{Lezama1999PRA}. 

\section{Correlations between cavity modes}
\label {sec:Correlations}

In this section, we show how the dilational mode of the membrane can be used to create two-mode squeezing between different cavity modes. We also show how this system can lead to an explicit observation of quantum back-action of the optical field on the mechanical oscillator \cite{Caves1980PRL}, also known as radiation pressure shot noise.

Let us now consider four optical modes $\hat{a}_i$, $i = 1,2,3,4$, where $\hat{a}_2$ ($\hat{a}_4$) and $\hat{a}_1$ ($\hat{a}_3$) differ in frequency by the mechanical oscillator frequency $\omega_\M$ such that scattering between these modes can occur through phonon creation or annihilation. In other words, we consider the Hamiltonian
\begin{eqnarray}
  \label{eq:4modesHam}
  H & = & \sum_{i=1}^{4} \hbar \omega_i \hat{a}^\dagger_i \hat{a}_i + \hbar \omega_\M \ccr \cde + \hbar \Gamma \left(\cde \, \hat{a}^\dagger_2 \hat{a}_{1} + \ccr \hat{a}^\dagger_1 \hat{a}_{2}  + \cde \, \hat{a}^\dagger_4 \hat{a}_{3} + \ccr \hat{a}^\dagger_3 \hat{a}_{4} \right) \ . \nonumber
\end{eqnarray}
Note that we do not require $\hat{a}_2$ and $\hat{a}_3$ to be adjacent modes, although they may be. It is also not necessary that $\omega_1,\omega_2  < \omega_3, \omega_4$. We imagine that both modes 2 and 3 are strongly driven at frequencies $\omega_{\mathrm{D},2} \sim \omega_2$ and $\omega_{\mathrm{D},3} \sim \omega_3$, respectively. This means that the mechanical oscillator's interaction with mode 1 is of the parametric amplifier type (Eq.~\eqref{eq:ParAmp}), whereas the interaction with mode 4 is of the beam splitter type (Eq.~\eqref{eq:beamSplitt}). This setup can lead to correlations between the modes 1 and 4. The reason is that if the mechanical dissipation is sufficiently small, one can adjust the drive strengths such that almost every phonon that is created by a downconversion of a photon from mode 2 to mode 1 will be destroyed by upconverting a photon from mode 3 to mode 4. In other words, the creation of a photon in mode 1 is almost always followed by a creation of a photon in mode 4. We now study how this can lead to measurable quantum correlations.

\subsection{Equations of motion and self-consistency}
\label{sec:selfConsistent}

We will assume that $\omega_{\mathrm{D},2} - \omega_1 - \omega_\M \ll \kappa, \gamma$ and $\omega_4 - \omega_{\mathrm{D},3} - \omega_\M \ll \kappa, \gamma$, which should not be a problem for $\kappa, \gamma$ in the kHz-MHz range. We transform the cavity and oscillator operators to frames rotating at their respective resonance frequencies, i.e.~$\hat{a}_i \rightarrow \ex^{-\im \omega_i t}\hat{a}_i$ and $\hat{c} \rightarrow \ex^{-\im \omega_\M t} \hat{c}$. Moreover, we can combine the vacuum noise entering cavity mode $i$ from all decay channels $j$ in the operator $\xihat_i(t) = \kappa^{-1/2} (\sqrt{\kappa_\mathrm{ex}} \, \xihat_{i,\mathrm{ex}} + \sum_j \sqrt{\kappa_j} \, \xihat_{i,j} )$. This leads to the equations of motion
\begin{eqnarray}
  \label{eq:EOM4Modes}
  \dot{\hat{a}}_1 & = & - \frac{\kappa}{2} \hat{a}_1 - \im  \alpha_2 \hat{c}^\dagger + \sqrt{\kappa} \, \xihat_1 \\ 
  \dot{\hat{a}}_4 & = & - \frac{\kappa}{2} \hat{a}_4 - \im  \alpha_3 \hat{c} + \sqrt{\kappa} \, \xihat_4  \nonumber \\ 
  \dot{\hat{c}} & = & - \frac{\gamma}{2} \hat{c} - \im  \left(\alpha_2 \hat{a}_1^\dagger + \alpha_3 \hat{a}_4 \right) + \sqrt{\gamma} \, \zetahat \nonumber 
\end{eqnarray}
for cavity modes 1 and 4 and the mechanical oscillator. The average amplitudes $\alpha_2 = \Gamma \bar{a}_2$ and $\alpha_3 = \Gamma \bar{a}_3$ with $\bar{a}_i = \langle \hat{a}_i \rangle$ have been assumed real and positive without loss of generality. These amplitudes should be determined self-consistently, which leads to the equations 
\begin{eqnarray}
  \label{eq:selfConsistent}
  \fl \qquad \bar{a}_2 & = & \bar{a}_{2,0} - \frac{\Gamma\alpha_2}{\left(\kappa + \gamma \right) \left(\kappa \gamma/4 + \alpha_3^2 - \alpha_2^2 \right)} \left[\gamma (n_\therm + 1) + \frac{\left(2 \kappa + \gamma \right) \alpha_3^2}{\kappa (\kappa + \gamma)/2 + \alpha_3^2 - \alpha_2^2} \right] \\
  \fl \qquad  \bar{a}_3 & = & \bar{a}_{3,0} - \frac{\Gamma \alpha_3}{\left(\kappa + \gamma \right) \left(\kappa \gamma/4 + \alpha_3^2 - \alpha_2^2 \right)} \left[\gamma n_\therm  + \frac{\left(2 \kappa + \gamma \right) \alpha_2^2}{\kappa (\kappa + \gamma)/2 + \alpha_3^2 - \alpha_2^2} \right] \ , \nonumber
\end{eqnarray}
where $\bar{a}_{i,0} = 2(\kappa_\mathrm{ex} P_i/\hbar \omega_{i} \kappa^2)^{1/2}$ is the amplitude in the absence of optomechanical coupling, and $P_2$ ($P_3$) is the power of the beam driving cavity mode $2$ ($3$). To avoid the mechanical oscillator becoming unstable \cite{Metzger2008PRL}, we focus on the case where $\bar{a}_3 \geq \bar{a}_2$. Also, we limit the discussion to the weak coupling limit $\Gamma \ll \gamma, \kappa$ and $\alpha_i$ not much greater than $\kappa$, relevant to the experiment we propose. In that case, the corrections to $\bar{a}_{i,0}$ on the right-hand side of Eqs.~\eqref{eq:selfConsistent} are small. Thus, it is a good approximation to replace $\alpha_i$ in these terms by $\alpha_{i,0} = \Gamma \bar{a}_{i,0}$. 

While the correction to $\alpha_2$ and $\alpha_3$ induced by the optomechanical coupling is negligibly small in the regime we study, we will see that Eqs.~\eqref{eq:selfConsistent} provide insight to the discussion about two-mode squeezing in the next section.

\subsection{Two-mode squeezing}

The fact that a photon creation in mode 4 almost always follows a photon creation in mode 1 can give rise to peculiar photon statistics, similar to what was studied in Ref.~\cite{Borkje2011PRL}. Here we study how it leads to two-mode squeezing. We define the internal cavity quadratures
\begin{eqnarray}
  \label{eq:IntramodeQuad}
  \hat{X}_{\theta_1}(t) & = & \frac{1}{2} \left( \hat{a}_1(t) \ex^{ - \im \theta_1} + \hat{a}^\dagger_1(t) \ex^{\im \theta_1} \right) \ , \\
  \hat{X}_{\theta_4} (t) & = & \frac{1}{2} \left( \hat{a}_4(t) \ex^{- \im \theta_4} + \hat{a}^\dagger_4(t) \ex^{\im  \theta_4} \right) \ . \nonumber
\end{eqnarray}
From these, we can define a general intermode quadrature
\begin{equation}
  \label{eq:IntermodeQuad}
  \hat{X}_{\theta_1,\theta_4} (t) = \frac{1}{\sqrt{2}} \left(\hat{X}_{\theta_1}(t) + \hat{X}_{\theta_4}(t) \right) \ .
\end{equation}
We note that in the absence of the mechanical oscillator, the vacuum fluctuations lead to the standard quantum limit $\langle \hat{X}^2_{\theta_1} \rangle = \langle \hat{X}^2_{\theta_4} \rangle = \langle \hat{X}^2_{\theta_1,\theta_4} \rangle = 1/4$ for the variance in these quadratures. 

Taking into account the coupling to the mechanical oscillator leads to squeezing in the quadrature $\hat{X}_{\theta_1,\theta_4}$ for appropriately chosen phases $\theta_1$ and $\theta_4$. Relative to the choice of defining $\bar{a}_2$ and $\bar{a}_3$ real, the optimal choice of phases is $\theta_1 + \theta_4 = 0$ modulo $2 \pi$. This results in the variance
\begin{eqnarray}
  \label{eq:IntSqueezingOptimal}
  \langle \hat{X}^2_{\theta_1,-\theta_1} \rangle  =  \frac{1}{4} \Bigg\{ 1 & + & \frac{(\alpha_3 - \alpha_2)}{(\kappa + \gamma) (\kappa \gamma/4 + \alpha_3^2 - \alpha_2^2)} \\ & \times & \Bigg[\gamma \Big((\alpha_3 - \alpha_2) n_\therm - \alpha_2 \Big) - \frac{(2 \kappa + \gamma) \alpha_2 \alpha_3 (\alpha_3 - \alpha_2) }{\kappa (\kappa + \gamma)/2 + \alpha_3^2 - \alpha_2^2} \Bigg] \Bigg\} \ . \nonumber
\end{eqnarray}
%where $\alpha_i  = \Gamma \sqrt{n_i}$. 
%We have assumed that the drive frequencies are chosen such that $\omega_{\mathrm{D},2} - \omega_1 - \omega_\M \ll \kappa, \gamma$ and $\omega_4 - \omega_{\mathrm{D},3} - \omega_\M \ll \kappa, \gamma$, which should not be a problem for $\kappa, \gamma$ in the kHz-MHz range. 
By using Eqs.~\eqref{eq:selfConsistent}, we can rewrite this result as
\begin{equation}
  \label{eq:IntSqueezingOptimal2}
  \langle \hat{X}^2_{\theta_1,-\theta_1} \rangle  =  \frac{1}{4} \Big\{ 1 + \left(\bar{a}_3 - \bar{a}_2\right) \Big[\left(\bar{a}_{3,0} - \bar{a}_3\right) - \left(\bar{a}_{2,0} - \bar{a}_2\right) \Big] \Big\} \ ,
\end{equation}
which shows that the two-mode squeezing occurs when the average amplitude deviation $\bar{a}_{i,0} - \bar{a}_i$ is larger for mode 2 than mode 3. From the discussion in Section \ref{sec:selfConsistent}, it is clear that we can replace $\alpha_i$ in Eq.~\eqref{eq:IntSqueezingOptimal} by $\alpha_{i,0}$ in the regime we are interested in where $\Gamma \ll \kappa, \gamma$, with $\alpha_{3,0} \geq \alpha_{2,0}$ and $\alpha_{i,0}$ not much greater than $\kappa$. 

\begin{figure}[htbp]
\begin{center} 
\includegraphics[width=.8\textwidth]{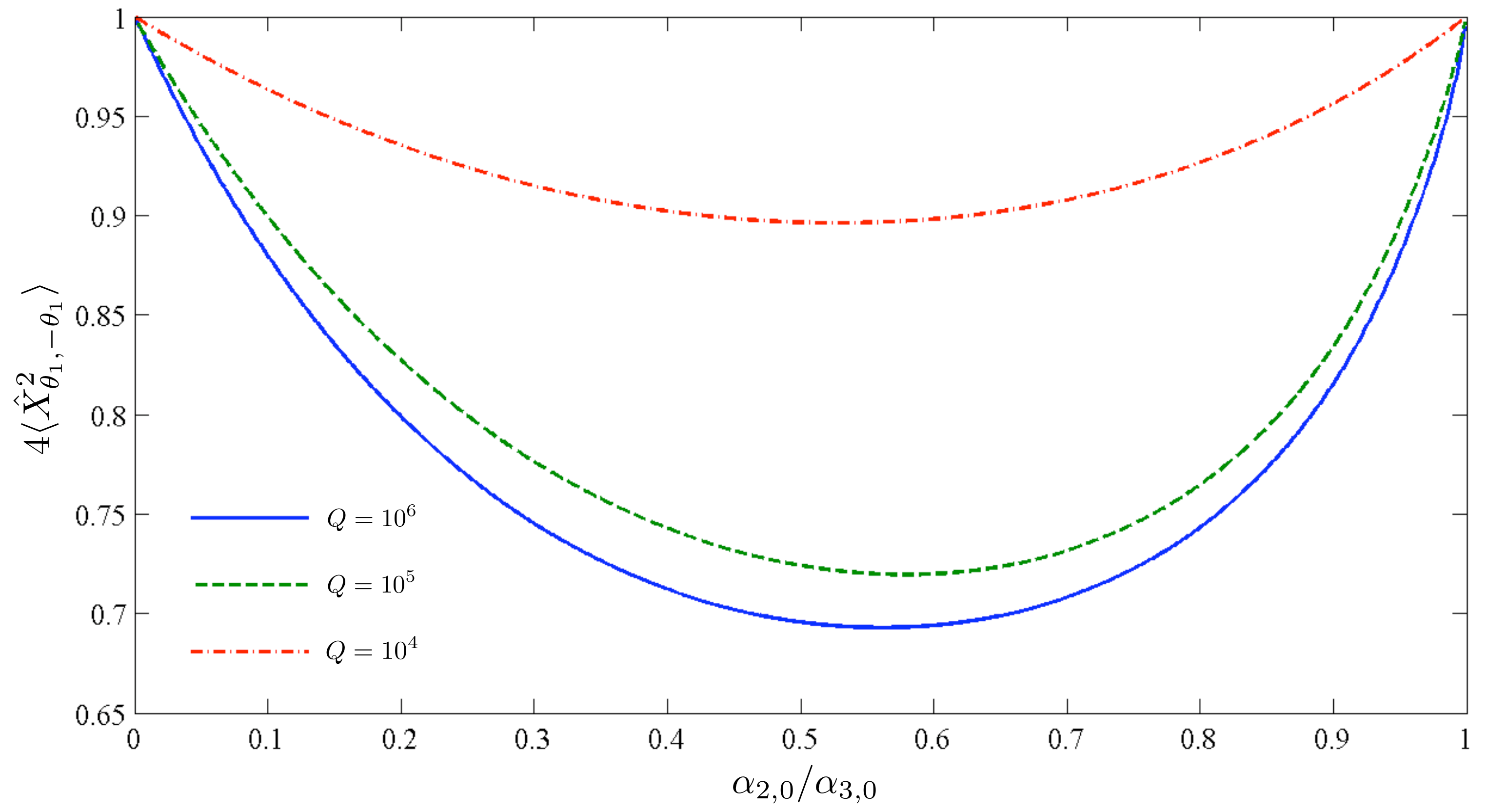}
\caption{The variance of the quadrature $\hat{X}_{\theta_1,-\theta_1}$ normalized to shot noise level, showing two-mode squeezing between modes 1 and 4. On the horizontal axis is drive strength at the blue sideband of mode 1 ($\alpha_{2,0}$) relative to drive strength at the red sideband of mode 4 ($\alpha_{3,0}$). We have assumed $\kappa/\alpha_{3,0} = 5/6$, $n_\therm = 0$, and plot for three levels of the mechanical linewidth, $\gamma = \kappa/10, \kappa$, and $10 \kappa$, corresponding to $Q = 10^4$, $10^5$, and $10^6$.}
\label{fig:TotTwoMode}
\end{center}
\end{figure}

The variance \eqref{eq:IntSqueezingOptimal} is shown in Figure \ref{fig:TotTwoMode} as a function of the ratio between the drives, $\alpha_{2,0}/\alpha_{3,0}$. The vertical axis is normalized to the shot noise level of the unsqueezed state. We use the parameters estimated from Section \ref{sec:estimation}, choosing $\kappa/\alpha_{3,0} = 5/6$ and $n_\therm = 0$. We plot for three values of the mechanical linewidth, $\gamma = 10 \kappa$, $\kappa$, and $\kappa/10$, corresponding to $Q = 10^4$, $10^5$, and $10^6$. We observe that the variance goes below the shot noise level, indicating two-mode squeezing. On the other hand, the individual variances $\langle \hat{X}^2_{\theta_1} \rangle , \langle \hat{X}^2_{\theta_4} \rangle \geq 1/4$, i.e.~there is no intramode squeezing.

The experimentally accessible quantities are the output quadratures
\begin{eqnarray}
  \label{eq:outputQuad}
  \hat{X}_{\theta_1, \mathrm{out}}(t) & = & \frac{1}{2} \left( \hat{a}_{1, \mathrm{out} }(t) \ex^{-\im \theta_1} + \hat{a}^\dagger_{1, \mathrm{out} }(t) \ex^{\im \theta_1} \right) \\
  \hat{X}_{\theta_4 , \mathrm{out}} (t) & = & \frac{1}{2} \left( \hat{a}_{4, \mathrm{out} }(t) \ex^{ -\im \theta_4} + \hat{a}^\dagger_{4, \mathrm{out} }(t) \ex^{\im \theta_4} \right) \ , 
\end{eqnarray}
where the output fields are $\hat{a}_{i, \mathrm{out} }(t) = \sqrt{\kappa_\mathrm{ex}} \hat{a}_i(t) - \xihat_{i,\mathrm{ex}}(t)$ and $\xihat_{i,\mathrm{ex}}(t)$ is the incoming vacuum noise at the input/output port where the quadratures are measured. These quadratures can be measured via homodyne photodetection, yielding access to the (two-mode) squeezing spectrum \cite{Carmichael1987JOptSocAmB,Drummond1990PRA}
\begin{equation}
  \label{eq:SqSpectrumDef}
  S_\mathrm{sq}[\omega] = 1 + 8 \int_0^\infty \du \tau \, \langle : \hat{X}_{\theta_1,\theta_4,\mathrm{out}} (0) \hat{X}_{\theta_1,\theta_4,\mathrm{out}} (\tau) : \rangle \cos(\omega \tau) 
\end{equation}
where $\hat{X}_{\theta_1,\theta_4,\mathrm{out}} = (\hat{X}_{\theta_1, \mathrm{out}} + \hat{X}_{\theta_4, \mathrm{out}})/\sqrt{2}$ and the colons indicate normal and time ordering. The spectrum is normalized in such a way that unity refers to shot noise level. We have neglected the fact that the two local oscillators used to measure $\hat{X}_{\theta_1, \mathrm{out}}$ and $\hat{X}_{\theta_4, \mathrm{out}}$ may have different strengths, and assumed a detector efficiency of unity. However, these are purely technical issues that can straightforwardly be taken into account \cite{Drummond1990PRA}. Exploiting the fact that $\xihat_{i,\mathrm{ex}}(t)$ is vacuum noise and that it cannot be correlated with the intracavity operators $\hat{a}_1(t'), \hat{a}_4(t')$ at earlier times $t' < t$ due to causality, the squeezing spectrum can be expressed in terms of the internal operators:
\begin{equation}
  \label{eq:SqSpectrumInternal}
  S_\mathrm{sq}[\omega] = 1 + 8 \, \kappa_\mathrm{ex} \int_0^\infty \du \tau \, \langle : \hat{X}_{\theta_1,\theta_4} (0) \hat{X}_{\theta_1,\theta_4} (\tau) : \rangle \cos(\omega \tau) \ .
\end{equation}
Focusing again on the optimally squeezed quadrature ($\theta_1 + \theta_4 = 0$ modulo $2 \pi$), we find
\begin{eqnarray}
  \label{eq:SqSpectrumExpression}
  S_\mathrm{sq}[\omega] = 1 & + & \kappa_\mathrm{ex} \left(\alpha_{3,0} - \alpha_{2,0} \right) |\chi_\C[\omega]|^2 |\chi_\M[\omega]|^2 \\ & \times & \Big\{ \gamma \big[ (\alpha_{3,0} - \alpha_{2,0}) n_\therm - \alpha_{2,0} \big] - \kappa \, \alpha_{2,0} \alpha_{3,0} \left(\alpha_{3,0} - \alpha_{2,0}\right) |\chi_\C[\omega]|^2 \Big\} \ . \nonumber
\end{eqnarray}
We have defined the susceptibilites
\begin{equation}
  \label{eq:susceptDef}
  \chi_\C[\omega] = \frac{1}{\kappa/2 - \im \omega} 
\end{equation}
and
\begin{equation}
  \chi_\M[\omega]  =  \frac{1}{\gamma/2 - \im \omega + (\alpha_3^2 - \alpha_2^2)\chi_\C[\omega]} \ .
\end{equation}
The validity of the expression \eqref{eq:SqSpectrumExpression} is also limited to the weak-coupling regime discussed above. 

\begin{figure}[htbp]
\begin{center} 
\includegraphics[width=.8\textwidth]{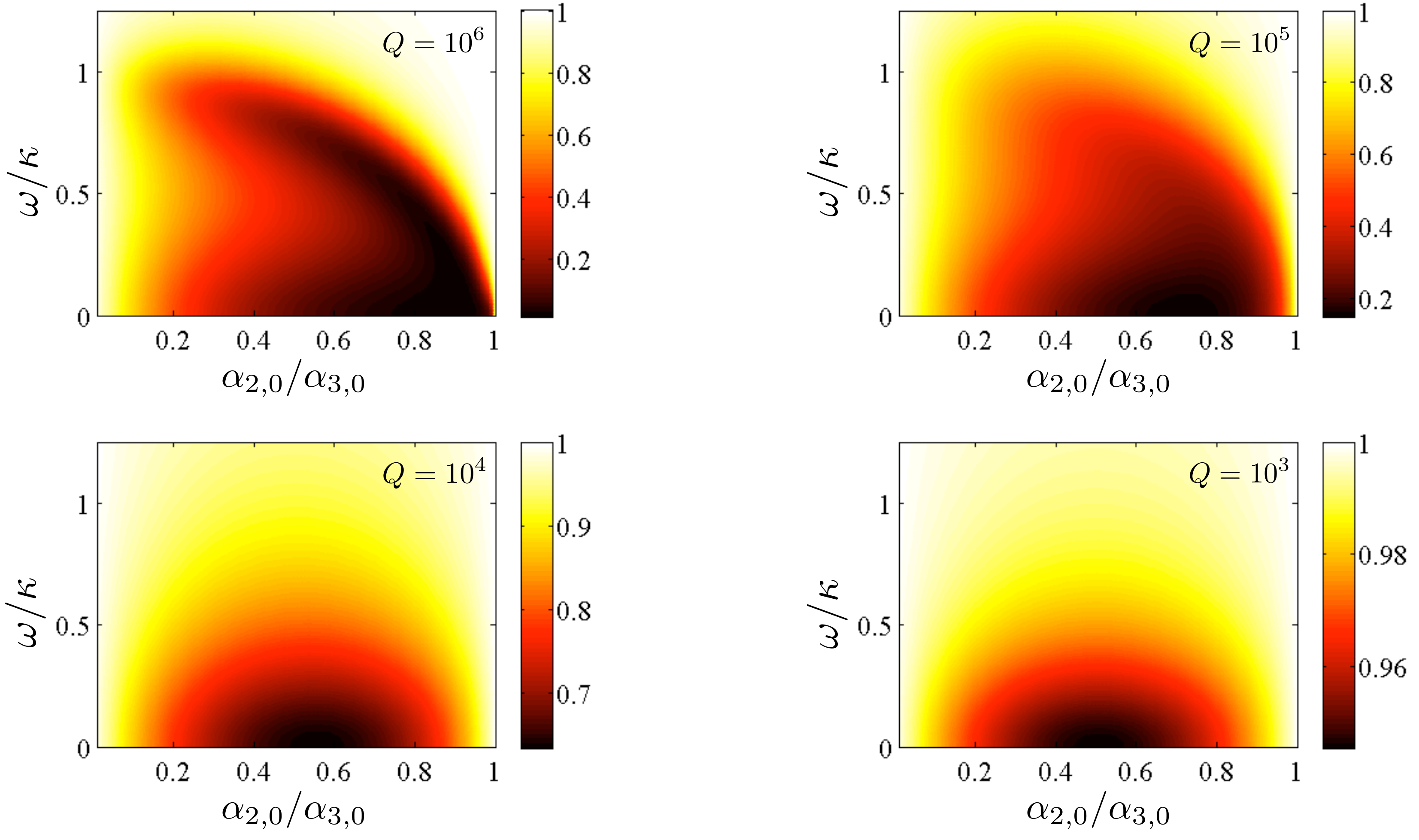}
\caption{The two-mode squeezing spectrum $S_\mathrm{sq}[\omega]$. On the vertical axis is $\alpha_{2,0}/\alpha_{3,0}$, i.e.~drive strength at the blue sideband of mode 1 relative to drive strength at the red sideband of mode 4. On the horizontal axis is frequency $\omega$ in units of $\kappa$. We have chosen $\kappa/\alpha_{3,0} = 5/6$ and $n_\therm = 0$. Top left: $\gamma/\kappa = 0.1$ ($Q = 10^6$). Top right: $\gamma/\kappa = 1$ ($Q = 10^5$). Bottom left: $\gamma/\kappa = 10$ ($Q = 10^4$). Bottom right: $\gamma/\kappa =  100$ ($Q = 10^3$).}
\label{fig:TwoMode}
\end{center}
\end{figure}
The output squeezing spectrum is plotted in Figure \ref{fig:TwoMode} as a function of frequency $\omega$ and the ratio between the drives, $\alpha_{2,0}/\alpha_{3,0}$. We have used the same parameters as in Fig.~\ref{fig:TotTwoMode}, and plot the spectrum for $Q = 10^3$, $10^4$, $10^5$, and $10^6$. We see that the spectrum goes well below shot noise level and that nearly maximal squeezing can be achieved at low frequencies $\omega < \kappa$ when the mechanical linewidth $\gamma < \alpha_{3,0}$, $\kappa$ and the ratio of the two drives is appropriately chosen. Even for a modest mechanical $Q = 10^3$, the squeezing spectrum goes several percent below the shot noise level.

It is clear from Eqs.~\eqref{eq:IntSqueezingOptimal} and \eqref{eq:SqSpectrumExpression} that the two-mode squeezing will diminish as the temperature and hence the equilibrium phonon number $n_\therm$ increases. For extremely large mechanical $Q$, the squeezing could nevertheless be observable at higher temperatures. For example, to make the variance \eqref{eq:IntSqueezingOptimal} go 1\% below the shot noise level at room temperature would require $Q > 3 \times 10^7$ with the parameters we have assumed.

\subsection{Radiation pressure shot noise}

Observation of the squeezing discussed above implies that the mechanical oscillator is susceptible to radiation pressure shot noise. We now discuss another way of using the two quadratures $\hat{X}_{\theta_1, \mathrm{out}}$ and $\hat{X}_{\theta_4, \mathrm{out}}$ to explicitly detect these correlations between the mechanical oscillator motion and photon shot noise. The method we propose is similar to the detection schemes studied in Refs.~\cite{Heidmann1997ApplPhysB,Borkje2010PRA}. However, the experimental setup we have studied here seems particularly appropriate for this kind of measurement, warranting a thorough analysis. 

As in the previous section, we focus on the model in Eq.~\eqref{eq:4modesHam} and imagine that modes $2$ and $3$ are strongly driven. Homodyne detection of $\hat{X}_{\theta_1, \mathrm{out}}(t)$ and $\hat{X}_{\theta_4, \mathrm{out}}(t)$ gives access to the cross-correlation spectrum
\begin{equation}
  \label{eq:crossCorr}
  S_\mathrm{cr}[\omega] = \frac{1}{2} \int^\infty_{-\infty} \du \tau \, \ex^{\im \omega \tau} \big\langle \big\{ \hat{X}_{\theta_1, \mathrm{out}}(\tau) \ ,  \ \hat{X}_{\theta_4, \mathrm{out}}(0) \big\} \big\rangle \ ,
\end{equation}
where $\{\cdot , \cdot\}$ denotes the anticommutator. This spectrum can be expressed in terms of two types of correlation functions. Some correlation functions are of the type $\langle \hat{c}^\dagger(\tau) \hat{c}(0) \rangle$, i.e.~mechanical oscillator autocorrelation functions. The others are of the type $\langle \xihat_i(\tau) \hat{c}(0) \rangle$, which are quantum optomechanical correlations between the vacuum noise of the electromagnetic field and the mechanical oscillator. The latter type is nonzero when the mechanical oscillator is susceptible to photon shot noise. In other words, these correlations functions are nonzero due to radiation pressure shot noise. 

\begin{figure}[htbp]
\begin{center} 
\includegraphics[width=.8\textwidth]{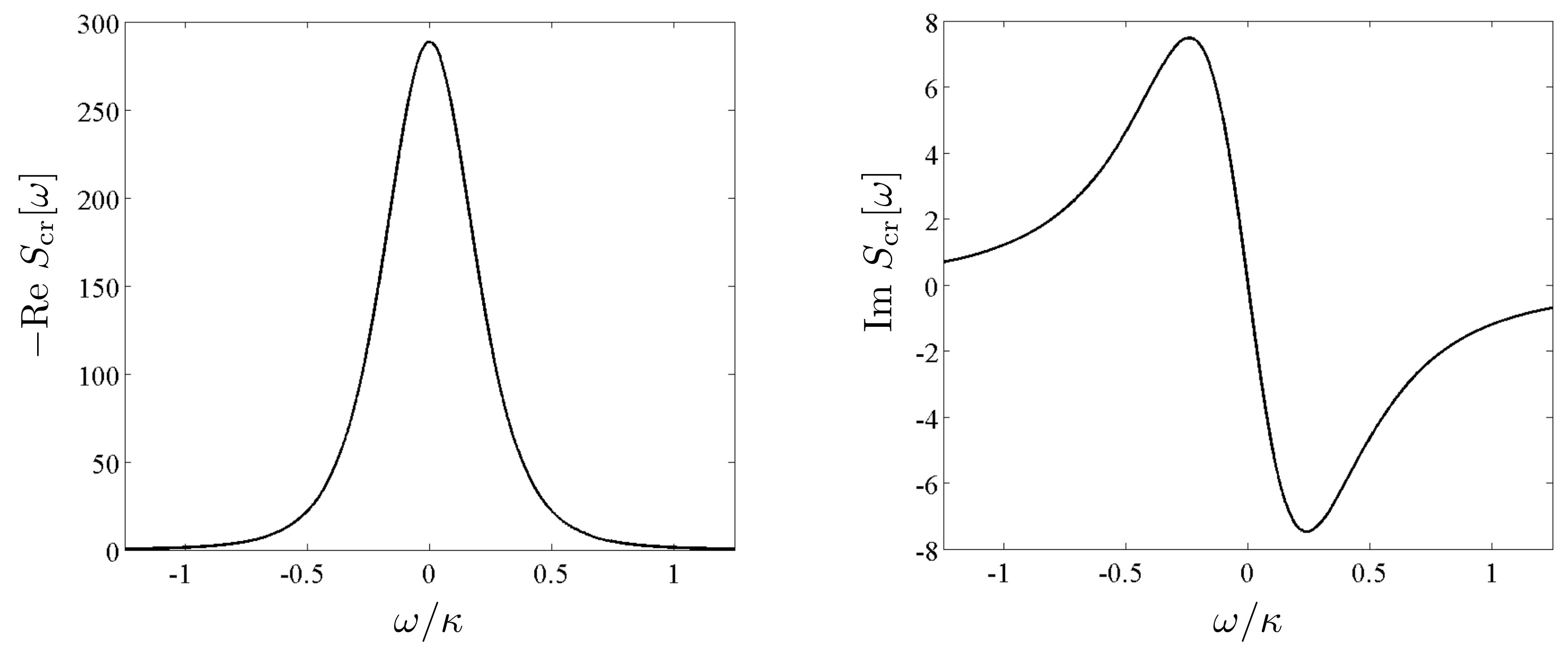}
\caption{The real (left) and imaginary (right) parts of the cross-correlation spectrum $S_\mathrm{cr}[\omega]$. We have used $\kappa/\alpha_{3,0} = 5/6$, $\alpha_{2,0}/\alpha_{3,0} = 1$, $\gamma/\kappa = 1$, and $n_\therm = 0$. The asymmetric imaginary part is a signature of radiation pressure shot noise.}
\label{fig:crosscorr}
\end{center}
\end{figure}

The cross-correlation spectrum becomes
\begin{equation}
  \label{eq:crossCorrCalc}
  S_\mathrm{cr}[\omega] = - \kappa_\mathrm{ex} \alpha_{2,0} \alpha_{3,0} |\chi_\C[\omega]|^2 |\chi_\M[\omega]|^2 \cos(\theta_1 + \theta_2) \big( R_\mathrm{cr}[\omega] + \im I_\mathrm{cr}[\omega] \big)\ , 
\end{equation}
where 
\begin{equation}
  \label{eq:realpart}
  R_\mathrm{cr}[\omega]  = \gamma \left(2 n_\mathrm{th} + 1 \right) + \kappa \left( \alpha_{2,0}^2 + \alpha_{3,0}^2 \right) |\chi_\C[\omega]|^2
\end{equation}
and
\begin{equation}
  \label{eq:imagpart}
I_\mathrm{cr}[\omega]  = - 2 \omega \Big[1 + \left(\alpha_{3,0}^2 - \alpha_{2,0}^2\right)|\chi_\C[\omega]|^2 \Big] \ .
\end{equation}
Note that the real part of $S_\mathrm{cr}[\omega]$ is even in frequency, whereas the imaginary part is odd. These functions are displayed in Figure \ref{fig:crosscorr} for the case of equal driving strengths, $\alpha_{2,0} = \alpha_{3,0}$. The real part comes from the oscillator autocorrelation functions, whereas the imaginary part is solely due to the correlations between the mechanical oscillator and the vacuum noise. Thus, measurement of an odd frequency imaginary part of the correlation function $S_\mathrm{cr}[\omega] $ is clear evidence of radiation pressure shot noise. 

In the calculation of $S_\mathrm{cr}[\omega]$, we have neglected technical laser noise, which can give rise to a similar asymmetry as the vacuum noise if it is not negligible at frequencies around $\omega_\M$ \cite{Borkje2010PRA}. However, for a mechanical frequency in the GHz range, neglect of the laser noise is a very realistic assumption. 

Finally, it is worth noting that the antisymmetric behaviour of the imaginary part $I_\mathrm{cr}[\omega] $ can in principle be measured at high temperatures. However, as the magnitude of the real part is proportional to temperature, control of phases so as to correctly identify the real and imaginary parts would have to be very accurate. Challenges of this type were discussed in more detail in Ref.~\cite{Borkje2010PRA}.

\section{Concluding remarks}

We have studied the optomechanical coupling between the thickness fluctuations of a thin dielectric membrane and an optical cavity field. For an appropriate choice of membrane thickness versus cavity length, the model reduces to one where the creation (annihilation) of a phonon causes scattering from an optical mode with higher (lower) frequency to one with lower (higher) frequency. We have derived the coupling constant and showed that realistic parameters suggest that optomechanical effects can be observable. We argued that this system can be used to amplify a narrow bandwidth optical signal with quantum limited added noise. Finally, we showed how one can create quantum correlations between two optical cavity modes at different frequencies. We calculated a two-mode squeezing spectrum and found considerable squeezing for realistic parameters. A method for clearly detecting radiation pressure shot noise was also presented. 

The physical system we discussed is already being actively studied experimentally in several labs, such that very small changes are necessary to attempt to observe these phenomena. A primary challenge is to detect the dilational motion, either through its thermal noise or its response to driving, and to determine the mechanical quality factor. If the latter is small, it would be an important technical challenge to find ways to increase it. With a sufficiently high quality factor, the possibility of observing quantum optomechanical effects with this high-frequency dilational mode appears feasible.

\section*{Acknowledgments}

We thank Jack Harris for numerous discussions, helpful suggestions, and a critical reading of the manuscript. KB acknowledges financial support from the Research Council of Norway under Grant No.~191576/V30, and The Danish Council for Independent Research under the Sapere Aude program. SMG acknowledges support from the National Science Foundation under Grant No.~DMR-1004406 and DMR-0653377.

\section*{References}

\bibliographystyle{unsrt}

%\bibliography{Dilational}

\end{document}